# Service Level Agreement (SLA) in Utility Computing Systems

**Linlin Wu and Rajkumar Buyya**
*Cloud Computing and Distributed Systems (CLOUDS) Laboratory*
*Department of Computer Science and Software Engineering*
*The University of Melbourne, Australia*
*Email: {linwu, raj}@csse.unimelb.edu.au*

**ABSTRACT**
In recent years, extensive research has been conducted in the area of Service Level Agreement (SLA) for utility computing systems. An SLA is a formal contract used to guarantee that consumers' service quality expectation can be achieved. In utility computing systems, the level of customer satisfaction is crucial, making SLAs significantly important in these environments. Fundamental issue is the management of SLAs, including SLA autonomy management or trade off among multiple Quality of Service (QoS) parameters. Many SLA languages and frameworks have been developed as solutions; however, there is no overall classification for these extensive works. Therefore, the aim of this chapter is to present a comprehensive survey of how SLAs are created, managed and used in utility computing environment. We discuss existing use cases from Grid and Cloud computing systems to identify the level of SLA realization in state-of-art systems and emerging challenges for future research.

**Keywords**: Service Level Agreement, Cloud computing, Grid computing, utility computing, SLA lifecycle, SLA Negotiation, SLA Management, SLA Languages and Frameworks.

## 1. INTRODUCTION

Utility computing (Yeo and Buyya 2006) delivers subscription-oriented computing services on demand similar to other utilities such as water, electricity, gas, and telephony. With this new service model, users no longer have to invest heavily on or maintain their own computing infrastructures, and they are not constrained to any specific computing service provider. Instead, they can outsource jobs to service providers and just pay for what they use. Utility computing has been increasingly adopted in many fields including science, engineering, and business (Youseff et. al. 2008). Grid, Cloud, and Service-oriented computing are some of the paradigms that have made delivery of computing as a utility. In these computing systems, different Quality of Service (QoS) parameters have to be guaranteed to satisfy user's request. A Service Level Agreement (SLA) is used as a formal contract between service provider and consumer to ensure service quality (Buco et. al. 2004).

*Figure 1* shows typical utility computing system architecture: User/Broker, SLA Management, Service Request Examiner, and Resource/Service Provider. **User or Broker** submits its requests via applications to the utility computing system, which includes bottom three layers. **Service Request Examiner** is responsible for Admission Control. **SLA Management** layer manages Resource Allocation. **Resource or Service Provider** offers resources or services.

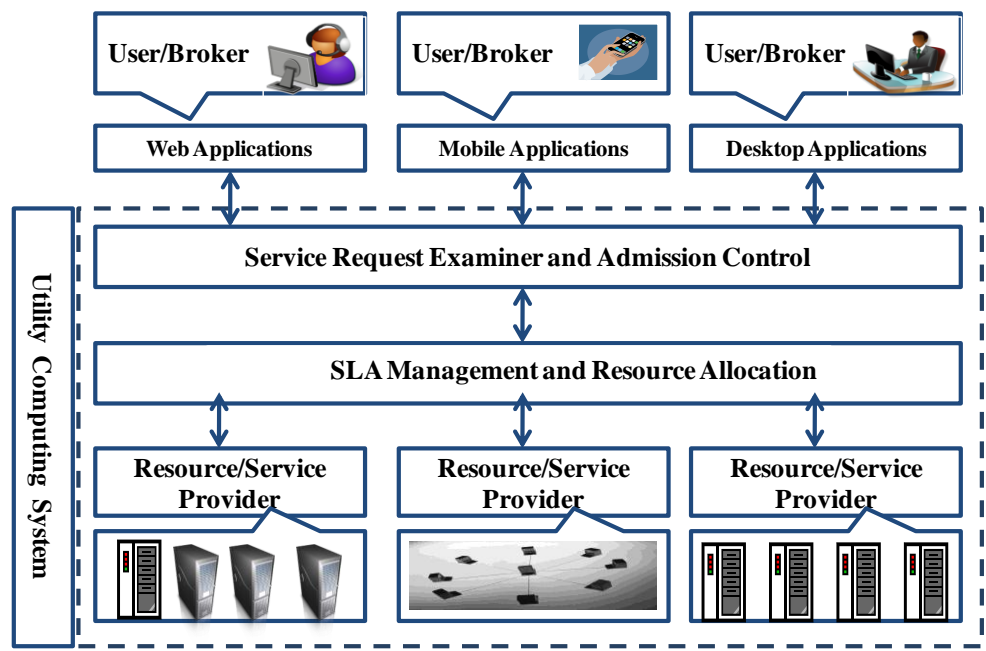

*Figure 1. A typical architectural view of utility computing system.*

In the above architecture, SLAs are used to identify parties who engage in the electronic business, computation, and outsourcing processes and to specify the minimum expectations and obligations that exist between parties (Buco et. al. 2004). The most concise SLA includes both general and technical specifications, including business parties, pricing policy, and properties of the resources required to process the service (Yeo et. al. 2006). According to Sun Microsystems Internet Data Center Group's report (2002), a good SLA sets boundaries and expectations of service provisioning and provides the following benefits:

- **Enhanced customer satisfaction level:** A clearly and concisely defined SLA increases the customer satisfaction level, as it helps providers to focus on the customer requirements and ensures that the effort is put on the right direction.

- **Improved Service Quality:** Each item in an SLA corresponds to a Key Performance Indicator (KPI) that specifies the customer service within an internal organisation.
- **Improved relationship between two parties**: A clear SLA indicates the reward and penalty policies of a service provision. The consumer can monitor services according to Service Level Objectives (SLO) specified in the SLA. Moreover, the precise contract helps parties to resolve conflicts more easily.

A clearly defined lifecycle is essential for effective realisation of an SLA. Ron, S. et. al. (2001) define SLA lifecycle in three high level phases, which are the creation phase, operation phase, and removal phase. Sun Microsystems Internet Data Center Group (2002) defines a practical SLA lifecycle in six steps, which are 'discover service providers', 'define SLA', 'establish agreement', 'monitor SLA violation', 'terminate SLA', and 'enforce penalties for violation'.

The realization of an SLA can be traced back to 1980s in telecommunication companies. Furthermore, the advent of Grid computing reinforces the necessity of using SLA (Yeo and Buyya 2006). Specifically, in service-oriented commercial Grid computing (Buyya et. al. 2001), resources are advertised and traded as services based on an SLA after users specify various levels of service required for processing their jobs (Rashid et. al. 2004). However, SLAs have to be monitored and assured properly (Sahai et. al. 2003). These works identified some challenges in SLA management, such as SLA violation control, which have been partially addressed by frameworks such as WS-Agreement (Andrieux et. al. 2007) and WSLA (Keller et. al. 2003). Still, in dynamic environments such as Clouds several challenges have to be addressed: automatic negotiation and dynamic SLA management according to environmental changes are the most important examples. .

Recently, Cloud computing has emerged as a new platform for delivering utility computing services. In Clouds, infrastructure, platform and application services are available on-demand and companies are able to access their business services and applications anywhere in the world whenever they need. In this environment, massively scalable systems are made available to end users as a service (Brandic 2009). In this scenario, where both request arrival rate and resources availability continuously vary, SLAs are used to ensure that service quality is kept at acceptable levels despite such dynamicity.

This chapter reveals key design factors and issues that are still significant in utility computing platforms such as Grids and Clouds. It provides insights for extending and reusing components of the existing SLA management frameworks and it aims to be a guide in designing and implementing enhanced SLA-oriented management systems. This work guides the design and implementation of enhanced SLA-oriented management systems.

The use cases selected for the chapter have been proposed recently (since 2004), and reflect the latest technological advances. The design concepts and architectures of these works are well-documented in publications to facilitate comprehensive investigation.

The rest of the chapter is organised as follows: Utility architecture and SLA foundational concepts are summarized in Section 2. In Section 3, the key challenges and solutions for SLA management are discussed. SLA use cases are proposed in Section 4. The ongoing works addressing some of the issues in current systems are presented in Section 5. Finally, the chapter concludes with the open challenges in SLA management in Section 6.

## 2. UTILITY ARCHITECTURE AND SLA FOUNDATIONS

In this section, initially, a typical utility computing architecture is presented. SLA definitions from different areas are summarized in Section 2.2. SLA components are described in Section 2.3. In Section 2.4, two types of SLA lifecycle are presented and compared.

### 2.1. Utility Architecture

The layered architecture of a typical utility computing system is shown in *Figure 2*. From top to bottom it is possible to identify four layers, a **User or Broker** submits its requests using various applications to the utility computing system, the **Service Request Examiner** is responsible for admission control, **SLA Management** balances workloads, and a **Resource or Service Provider** offers resources or services. Users or Brokers, who act on users' behalf, submit their service requests and applications, from anywhere in the world, to be processed by utility computing systems. When a service request is submitted, the Service Request Examiner uses Admission Control mechanism to interpret its QoS requirements before determining whether to accept or reject it. Thus, it ensures that there is no overloading of resources whereby many service requests cannot be fulfilled successfully due to limited availability of resources/services. Then, the Service Request Examiner interacts with the SLA Management to decide whether to accept or reject the request.

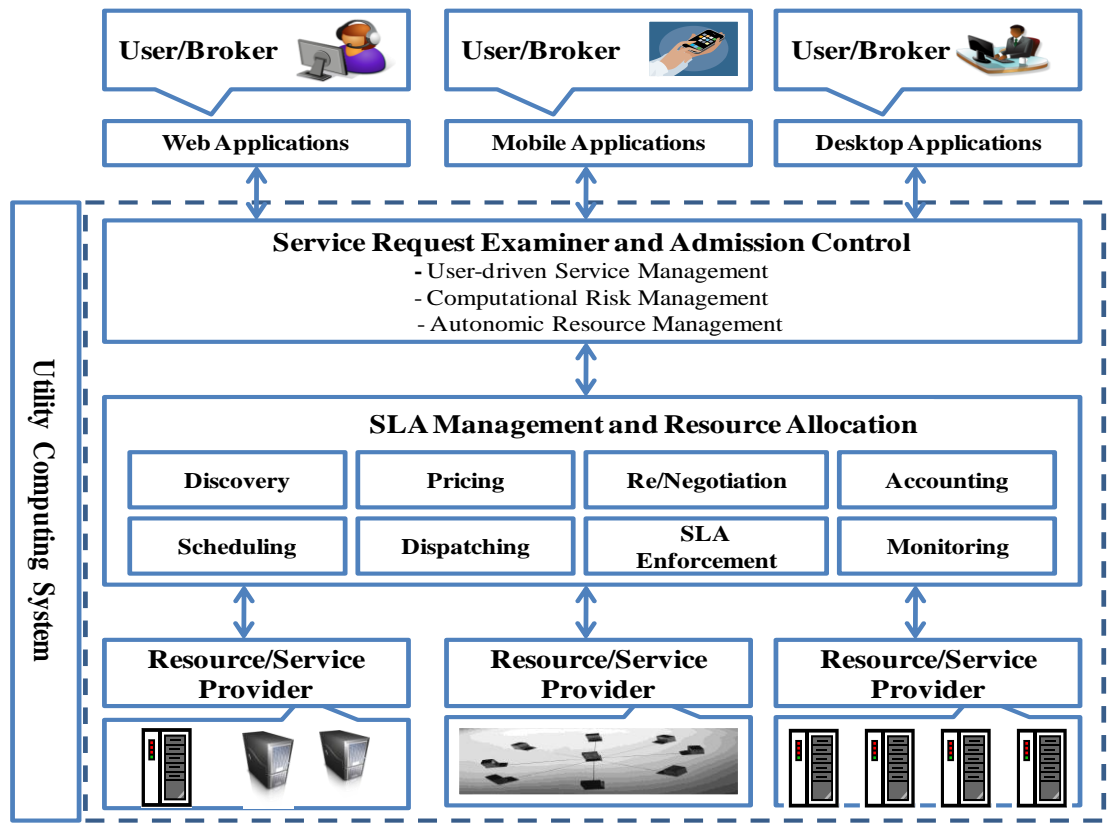

*Figure 2. SLA-Oriented Utility Computing System Architecture*

The SLA Management component is responsible for resource allocation and consists of several components: Discovery, Negotiation/Renegotiation, Pricing, Scheduling, Monitoring, SLA Enforcement, Dispatching and Accounting. The Discovery component is responsible for discovering service providers that can satisfy user requirements. In order to define mutually agreed terms between parties, it is common to put in place price negotiation mechanisms or to rely on quality metrics. The Pricing mechanism decides how service requests are charged. Pricing serves as a basis for managing supply and demand of computing resources within the utility computing system, and facilitates in prioritizing resource allocations. Once the negotiation process is completed, the Scheduling mechanism uses algorithms or policies to decide how to map requests to resource providers. Then the Dispatching mechanism starts the execution of accepted service requests on allocated resources.

The Monitoring component consists of a Resource Monitoring mechanism and a Service Request Monitoring mechanism. The Resource Monitoring mechanism keeps track of the availability of Resource Providers and their resource entitlements. On the other hand, the Service Request Monitoring mechanism keeps track of the execution progress of service requests. The SLA enforcement mechanism manages violation of contract terms during the execution. Due to the SLA violation, sometimes Renegotiation is needed in order to keep ongoing trading. The Accounting mechanism maintains the actual usage of resources by requests so that the final cost can be computed and charged to the users. At the bottom of the architecture, there exists a Resource/Service Provider that comprises multiple services such as computing services, storage services and software services in order to meet service demands.

## 2.2. SLA Definitions

Dinesh et. al. (2004) define an SLA as: "*An explicit statement of expectations and obligations that exist in a business relationship between two organizations: the service provider and customer*". Since SLA has been used since 1980s in a variety of areas, most of the available definitions are contextual and vary from area to area. Some of the main SLA definitions in Information Technology related areas are summarised in *Table 1*.

Table 1: Summary of SLA definitions classified by the area.

| Area | Definition | Source |
|------|-----------|--------|
| Web Services | "*SLA is an agreement used to guarantee web service delivery. It defines the understanding and expectations from service provider and service consumer*". | HP Lab (Jin et. al. 2002) |
| Networking | "*An SLA is a contract between a network service provider and a customer that specifies, usually in measurable terms, what services the network service provider will supply and what penalties will assess if the service provider can not meet the established goals*". | Research Project |
| Internet | "*SLA constructed the legal foundation for the service delivery. All parties involved are users of SLA. Service consumer uses SLA as a legally binding description of what provider promised to provide. The service provider uses it to have a definite, binding record of what is to be delivered*". | Internet NG (Ron et. al.2001) |
| Data Center Management | "*SLA is a formal agreement to promise what is possible to provide and provide what is promised*". | Sun Microsystems Internet Data Center group (2002) |

## 2.3. SLA Components

An SLA defines the delivery ability of a provider, the performance target of consumers' requirement, the scope of guaranteed availability, and the measurement and reporting mechanisms (Rick, 2002).

Jin et. al. (2002) provided a comprehensive description of the SLA components, including: (*Figure 3*):

- **Purpose**: Objectives to achieve by using an SLA.
- **Restrictions**: Necessary steps or actions that need to be taken to ensure that the requested level of services are provided.
- **Validity period:** SLA working time period.
- **Scope**:  Services that will be delivered to the consumers, and services that will not be covered in the SLA.
- **Parties**: Any involved organizations or individuals involved and their roles (e.g. provider and consumer).
- **Service-level objectives (SLO)**: Levels of services which both parties agree on. Some service level indicators such as availability, performance, and reliability are used.
- **Penalties:** If delivered service does not achieve SLOs or is below the performance measurement, some penalties will occur.
- **Optional services**: Services that are not mandatory but might be required.
- **Administration**: Processes that are used to guarantee the achievement of SLOs and the related organizational responsibilities for controlling these processes.

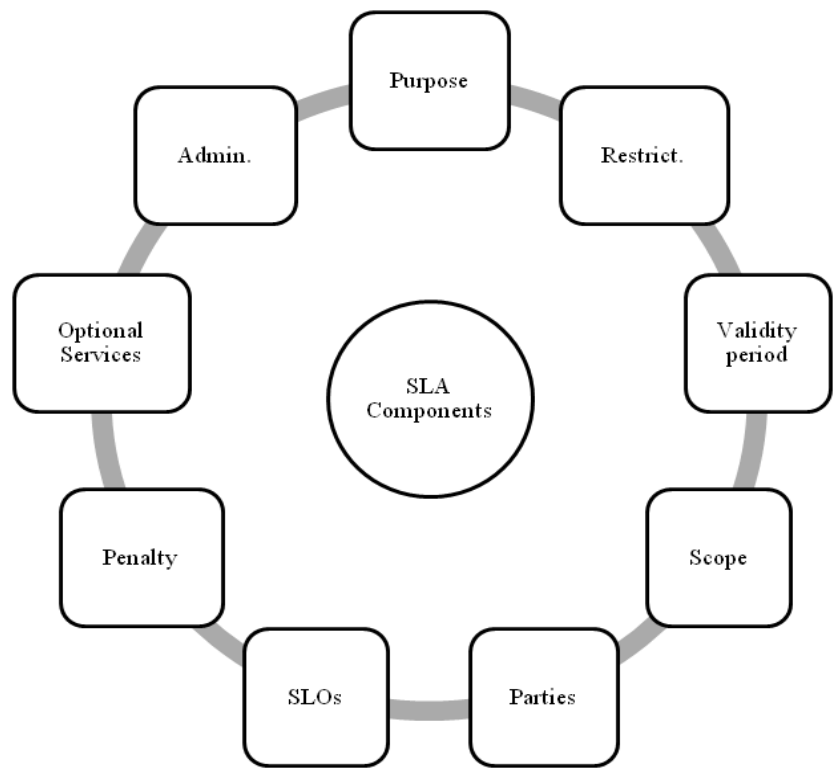

*Figure 3. SLA Components.*

## 2.4. SLA Lifecycle

Ron et. al. (2001) define the SLA life cycle in three phases (*Figure 4*). Firstly, the **creation phase,** in which the customers find service provider who matches their service requirements. Secondly, the **operation phase,** in which a customer has read-only access to the SLA. Thirdly, the **removal phase,** in which SLA is terminated and all associated configuration information is removed from the service systems.

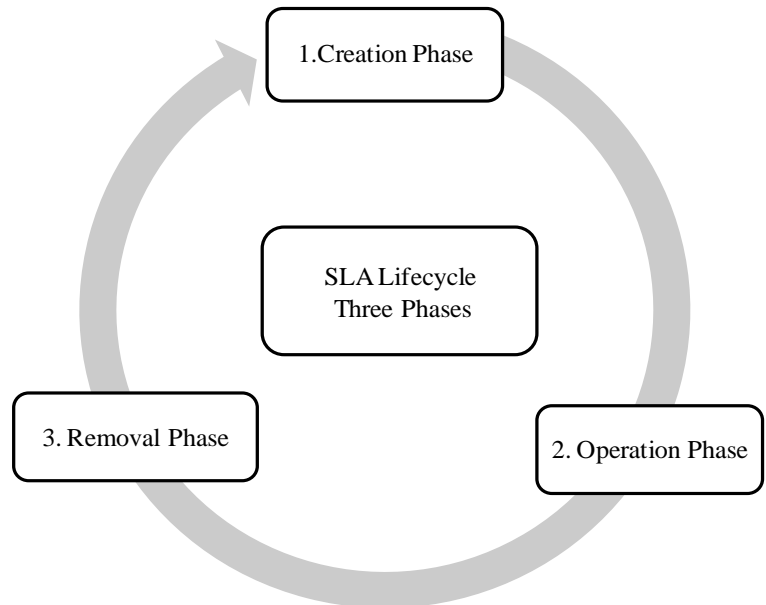

*Figure 4. SLA high level lifecycle phases, according to the description of Ron et. al. (2001).*

A more detailed life cycle has been characterized by the Sun Microsystems Internet Data Center Group (2002) , which includes six steps for the SLA life cycle: the first step is **'discover - service providers'**, in where service providers are located according to consumer's requirements. The second step is **'define – SLA',** which includes definition of services, parties, penalty policies and QoS parameters. In this step it is possible to negotiate between parties to reach a mutual agreement. The third step is **'establish – agreement',** in which an SLA template is established and filled in by specific agreement, and parties are starting to commit to the agreement. The fourth step is **'monitor – SLA violation'**, in which the provider's delivery performance is measured against to the contract. The fifth step is **'terminate – SLA'**, in which SLA terminates due to timeout or any party's violation. The sixth step is **'enforce - penalties for SLA violation'**, if there is any party violating contract terms, the corresponding penalty clauses are invoked and executed. These steps are illustrated in *Figure 5*.

The mapping between three high level phases and six steps of SLA lifecycle is shown in *Table 2*. The 'creation' phase of three phase lifecycle maps to the first three steps of the other lifecycle. In addition, the 'operation' phase of three phase lifecycle is the same as the fourth step of the other lifecycle. The rest of the phases and steps map to each other.

*Table 2: Mapping between two types of SLA lifecycle.*

| Three Phases | Six Steps |
|---|---|
| 1.  Creation Phase | 1.  Discover Service Provider |
|  | 2.  Define SLA |
|  | 3.  Establish Agreement |
| 2.  Operation Phase | 4.  Monitor SLA Violation |
| 3.  Removal Phase | 5.  Terminate SLA |
|  | 6.  Enforce Penalties for SLA Violation |

The six steps SLA lifecycle is more reasonable and provides detailed fine grain information, because it includes important processes, such as re/negotiation and violation control. During the service negotiation or renegotiation, a consumer exchanges a number of contract messages with a provider in order to reach a mutual agreement. The result of these processes leads to a new SLA (Youseff et. al. 2008). In six steps lifecycle, steps 2 and 3 map to these processes. However, the three phase's lifecycle does not include them. Furthermore, the 'Enforce Penalties for SLA violation' phase is important because it motivates parties adhere to follow the contract. We believe that the six steps formalization of the SLA life cycle provides a better characterization of the phenomenon and from here onwards we will refer to this as SLA life cycle.

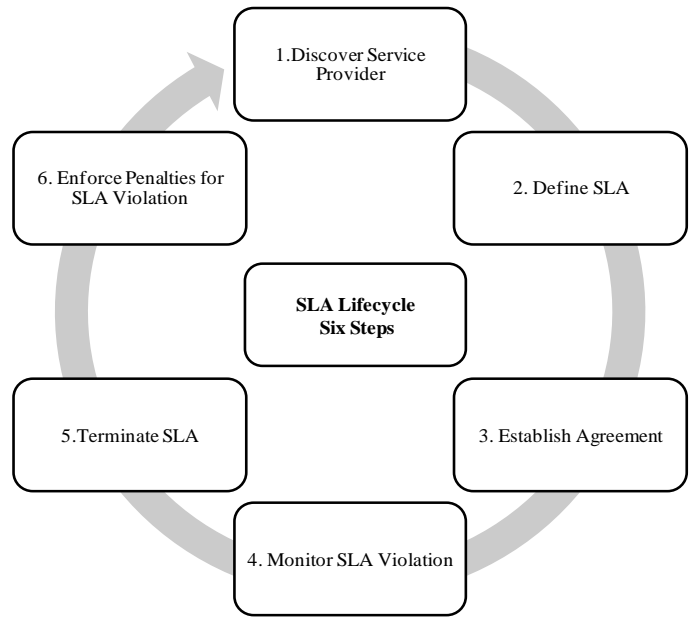

*Figure 5. SLA life cycle six steps, as defined by Sun Microsystems Internet Data Center Group (2002).*

## 3. SLA IN UTILITY COMPUTING SYSTEMS

As highlighted by Patterson (Patterson, 2008), there are many challenges involved in developing software for a million users to use as a service via a data center as compared to distributing software for a million users to run on their individual personal computers. Using SLAs to define service parameters that are required by users, the service provider knows how users value their service requests, hence it provides feedback mechanisms to encourage and discourage service request submissions. In particular, utility models are essential to balance the supply and the demand of computing resources by selectively accepting and fulfilling limited service requests out of many competing service requests submitted.

However, in the case of service providers making available a commercial offer to enable crucial business operations of companies, there are other critical QoS parameters to be considered in a service request, such as reliability and trust/security. In particular, QoS requirements cannot be static and need to be dynamically updated over time due to continuing changes in business operations and operating environments. In short, there should be greater importance on customers since they pay for accessing services. Therefore, the emphasis of this section is to describe SLA management in utility computing systems.

### 3.1. SLA Management in Utility Computing Systems

SLA management includes several challenges and in this section we will discuss them as part of the steps of the SLA life cycle.

### 3.1.1 Discover - Service Provider

In current utility computing environments, especially Grid and Cloud, it is important to locate resources that can satisfy consumers' requirement efficiently and optimally (Gong et. al. 2003). Such computing environments contain a large collection of different types of resources, which are distributed worldwide. These resources are owned and operated by various providers with heterogeneous administrative policies. Resources or services can join and leave a computing environment at anytime. Therefore, their status changes dynamically and unpredictably. Solutions for service provider discovery problems must efficiently deal with scalability, dynamic changes, heterogeneity and autonomous administration.

### 3.1.2. Define - SLA

Once service providers have been discovered, it is necessary to identify the various elements of an SLA that will be signed by agreeing metrics. These elements are called service terms and include QoS parameters, the delivery ability of the provider, the performance target of diversity components of user's workloads, the bounds of guaranteed availability and performance, the measurement and reporting mechanisms, the cost of the service, the data set for renegotiation, and the penalty terms for SLA violation. In this stage of the SLA lifecycle, measurement metrics and definition of each of these elements is done by a negotiation process between both parties (Blythe et. al. 2004) (Chu et. al. 2002).

Other challanges are related tothe negotiation process. Firstly, parties may use different negotiation protocols or they may not have the common definition of the same service (Brandic et. al. 2008). Secondly, service descriptions, in an SLA, must be defined unambiguously and be contextually specified by the means of its domain and actor. Therefore, an SLA language must allow the parameterisation of service description (Loyall et. al. 1998). Moreover it should allow a high degree of flexibility and enable a precise formalisation of what a service guarantee means. Another aspect is how to keep SLA definition consistent throughout the entire SLA lifecycle.

### 3.1.3. Establish - Agreement

In this step an SLA template is constructed. A template has to include all aspects of SLA components. In utility computing environments, to facilitate dynamic, versatile, and adaptive IT infrastructures, utility computing systems have to promply react to environmental changes, software failures, and other events which may influence the system's behavior. Therefore, how to manage SLA-oriented adaptive systems, which exploit self-renegotiation after system failure, becomes an open issue (Brandic et. al. 2009). Although most of the works recognise SLA negotiation as a key aspect of SLA managemet, recent works only provide little insight on how negotiation (especially automated negotiation) can be realised. In addition, it is difficult to reflect the quality aspects of SLA components in a template.

### 3.1.4. Monitor - SLA Violation

SLA violation monitoring begins once an agreement has been established. It plays a critical role in determining whether SLOs are achieved or violated. There are three main concerns. Firstly, which party should be in charge of this process. Secondly, how fairness can be assured between parties. Thirdly, how the boundaries of SLA violation are defined.

SLA violation means '*un-fulfillment*' of service agreement. According to the Principles of European Contract Law, the term '*un-fulfillment'* is defined as defective performance (parameter monitored at lower level than agreed), late performance (service delivered at the appropriate level but with unjustified delays), and no performance (service not provided at all). There are three broad provisioning categories based on the above definition (Rana et. al. 2008). '*All-or-Nothing*' provisioning, characterizes the case in which all SLOs must be satisfied or delivered by the provider. '*Partial*' provisioning identifies some SLOs as mandatory ones, and must be met for the successful service delivery by both parties. '*Weighted Partial*' provisioning, is the case in which the "*provision of a service meets SLO if it has a weight greater than a threshold (defined by the client)*" (Rana et. al. 2008). '*All-or-Nothing*' provisioning is used in most cases of SLA violation monitoring, because violation leads to complete failure and negotiation to create a new SLA. An SLA contains mandatory SLOs that must be delivered by the provider. Hence, in '*Partial*' provisioning, all parties assign these SLOs the highest priority to reduce violation risk. How much the SLO affects the '*Business Value*' a measure of the importance of a particular SLO term. The more important the violated SLO, the more difficult it is to renegotiate the SLA, because any party does not want to lose their competitive advantages in the market.

### 3.1.5. Terminate - SLA

In terminating an SLA, a key aspect is to decide when it should be terminated, and once decided, all associated configuration information is removed from the service systems. If the termination is due to an SLA violation, two questions need to be answered, who is the party that triggered this activity and what are the consequences of it.

### 3.1.6. Enforce Penalties for SLA Violation

In order to enforce penalties for SLA violation, penalty clauses are need to be defined. In utility computing systems, where consumers and providers are globally distributed, the penalty clauses work differently in various countries.

This leads to two problems, which particular clause should be used and whether it is fair for both sides. Moreover, due to the different types of violations, the penalty clauses need to be comprehensive. Recently, some works used the linear model for penalty enforcement of SLA violations in simple contexts (Lee et. al., 2010) (Yeo et. al., 2008). The linear model exhibits a poor performance, thus, the selection of these best models for SLA violation penalty clauses enforcement is still an open problem.

## 3.2. Solutions for SLA Management in Utility Computing Systems

This section introduces solutions for the problems presented in the previous section. Six SLA management languages and frameworks are analyzed, because they can be used as solutions in multiple steps of SLA lifecycle.

### 3.2.1. SLA Management Frameworks and Languages

SLA can be represented by specialized languages for easing SLA preparation, automating SLA negotiation, adapting services automatically according to SLA terms, and reasoning about their composition. In this section we introduce six languages for SLA specification and management. Among them, the WS-Agreement and Web Service Level Agreement (WSLA) are the most popular and widely used in research and industry. The comparison among all of these languages is shown in *Table 3*.

**Bilateral Protocol:** (Srikumar et. al. 2008) presented a negotiation mechanism for advanced resource reservation. It is a protocol for negotiating SLAs based on Rubinsteins Alternating Offers protocol for bargaining between parties. Any party is allowed to modify the proposal in order to reach a mutually-agreed contract. The authors implemented this protocol by using the Gridbus Broker on the customer's side and Aneka on the provider's side. Web services enable platform independence, and are therefore used to communicate between consumers and providers because the Gridbus Broker is implemented in Java, and Aneka is a .Net based enterprise Grid. The advantage of these high level languages is that they are object oriented and web services enable semantic definition. Thus, this protocol supports SLA component reuse, and type and semantic definition.

**WS-Agreement:** Open Grid Forum (OGF) has defined a standard for the creation and the specification of SLAs called Web Services Agreement Specification (WS-Agreement) (Andrieux et. al. 2007). It is a language and a protocol for establishing, negotiating, and managing agreements on the usage of services at runtime between providers and consumers. It uses an Extensible Markup Language (XML) based language for specifying the nature of an agreement template, which facilitates discovery of compatible providers. Its interaction is based on request and response. Moreover, it helps parties in exposing their status, so SLA violation can be dynamically managed and verified. Originally the language did not support negotiation and currently it has been complemented. WS-Agreement Negotiation, which lies on the top of WS-Agreement and describes the re/negotiation of the SLA. Its main feature is the robust signaling protocol for the negotiation.

**Web Service Level Agreement (WSLA)**: WSLA (Keller et. al. 2003) is a framework developed by IBM to specify and monitor SLA for Web Services. It provides a formal XML schema based language to express SLAs, and architecture to interpret this language at runtime. It can measure, and monitor QoS parameters and report violations to the parties. It separates monitoring clauses from contractual terms for outsourcing purposes. It provides the capability to create new metrics over existing metrics to implement multiple QoS parameters (Keller et. al. 2003). However, the semantic of metrics is not formally defined, hence, there are limitations for the creation of new terms based on existing terms.

**WSOL**: Web Service Offerings Language (WSOL) defines a syntax for service offers' interaction (Sakellariou et. al. 2005). It provides template instantiation and reuse of definitions (Buyya et. al. 2009). WSOL and WSLA support definition of management information and actions, such as violation notifications. However, they are not defined by a formal semantic. WSOL and QML (Quality of Service Management Language) support type systems allowing the same SLA to be described either in abstract or specific values to create a new SLA. The generalisation relationships between SLAs facilitate definitions of SLA types.

**SLAng**: Skeneet et. al. (2004) propose Service Level Agreement Language (SLAng), which uses XML to define SLAs. It is motivated by the fact that federated distributed systems must manage the quality of all aspects of their deployment. SLAng is different from other languages and frameworks. Firstly, it defines an SLA vocabulary for internet services. Secondly, its structure is based on the specific industry requirement, aiming to provide usable terms. Thirdly, it is modeled using Unified Markup Language (UML) and defined according to the behavior of services and consumers involved in service usage, unlike other languages, such as WSLA and WSOL, where QoS definition is based on metrics. Moreover, it supports third party monitoring schemes. However, it lacks of the ability to define management information, such as associated financial terms. Thus, it is not suitable for commercial computing environments.

**QML:** QML (Frolund et. al. 1998) defines a type system for SLAs, allowing users to define their own dimension types. However, it does not support extension of individual defined metrics because the exchange of SLAs between parties requires a common understanding of metrics. QML defines semantic for both its type system and its notion of SLA conformance.

**QuO:** Quality Objects (QuO) is a CORBA specific framework for QoS adaption based on proxies (Loyall et. al. 1998). It includes a quality description language used for describing QoS parameters, adaptations and notifications. QuO properties are the response of invoking instrumentation methods on remote objects. Like WSLA, no formal constraints are placed on the implementation of these methods.

### 3.2.2. Discover - Service Provider

In the Grid computing community, Fitzgerald (1997) introduced the Monitoring and Discovery System, Gong et. al. (2003) proposed the VEGA Grid Project and also relevant is the work of Iamnitchi et. al. (2001).

Monitoring and Discovery System (MDS) is the information service described in the Globus project (Fitzgerald 1997). In its architecture, Lightweight Directory Access Protocol (LDAP) is used as directory service, and information stored in information servers are organised in tree topology. In utility computing systems, resources' availability and capability are dynamic in nature. However, in MDS, the relationship between information and information servers is static. In addition, service provider's information is frequently updated in these dynamic changing environments, whilst LDAP is not designed for writing and updating information.

VEGA Infrastructure for Resource Discovery (VIRD) follows three-level hierarchy architecture. The top level is a backbone, which is responsible for the inter-domain resource discovery and consists of Border Grid Resource Name Servers (BGRNS). The second level consists of several domains and each domain consists of Grid Resource Name Servers (GRNS). The third level includes all clients and resource providers. There is no central control in this architecture, thus resource providers register themselves to GRNS server within a domain. When clients submit requests, GRNS respond to them with requested resources. The limitation of this architecture is that it only focuses on the issue of scalability and dynamic environmental changes but not on heterogeneity and autonomous administration.

Iamnitchi et. al. (2001) propose a resource discovery framework using peer-to-peer (P2P) technologies in Grids. P2P architecture is fully distributed and all the nodes are equivalent. However, one major limitation of their work is that every node has little knowledge about resources distribution and their status. Specifically, when there is a large number of resource types or the work-set is very large, the opportunity for inaccurate results increases, because the framework is not able to use historical data to accurately discover resources.

### 3.2.3. Define - SLA and Establish - Agreement

'Define – SLA' and 'Establish – Agreement' are two dependent steps, and SLA languages facilitate their development. For example, WSLA and WS-Agreement are the most widely used languages in these steps. Creation and Monitoring of Agreements (CREMONA) is a WS-Agreement framework implemented by IBM (Dan et. al. 2004). It proposes a Commitment Agreement and architecture for the WS-Agreement. All of these agreements are normal WS-Agreements, following a certain naming convention. This protocol basically aims at solving problems related to the creation of agreements on multiple sites. However, it is unable to solve limitations when service providers and consumers have different standards, policies, and languages during negotiations. For example, if a consumer uses WSLA but a provider uses WS-Agreement, the interaction is actually not possible. In order to solve this, Brandio et. al. (2008) proposed a Meta-Negotiation Architecture for SLA-Aware Grid Services based on meta-negotiation documents. These documents record supported protocols, document languages, and the prerequisites for starting negotiations and establishing agreements for all participants.

SLA-oriented Resource Management Systems (RMS) have been developed for addressing negotiation problems in Grids, for example, Wurman et. al. (1998) state a set of auction parameters and a price-based negotiation platform, which serves as an auction server for humans and software agents. Nevertheless, their solution only supports one-dimensional auction (only focus on price), but not multiple-dimensional auctions, which are important in utility computing environments.

Table 3. *Comparison of SLA Management frameworks and Languages.*

| Name | Type | Domain | Dynamic Establish / Management | Negotiation | Metrics | Define Management Actions | Support Reuse | Provide Type Systems | Define Semantic | Cope with SLA life-cycle |
|------|------|--------|-------------------------------|-------------|---------|--------------------------|---------------|----------------------|-----------------|--------------------------|
| Bilateral Protocol | Java, .Net and Web Service based protocol | Originally for resource reservation in Grids. | Yes | Yes | Yes | Yes | Yes. | Yes | Support by Web Service. | Step 1 to Step 4. |
| WS-Agreement | XML language; Framework; A protocol | Any domain | Establish and manage dynamically | Re/negotiation with WS-Agreement Negotiation | Do not define specification of metrics associated with agreement parameters. | Yes | Yes | Yes | Not formally defined | Step 1 to step 6 |
| WSLA | Provide language; Framework; runtime architecture | Originally for Web services | Establish and manage dynamically | Re/negotiation. | Allows creation of new metrics | Yes | Yes | NA | Not formally defined | Step 1 to step 6 |
| QML | language | Any Domain | Yes | Yes | Allows creation of new metrics | Yes | Yes | Yes, allows definition of new type systems | Yes | Step 1 to step 4 |
| WSOL | XML | Originally for Web Services | Yes | Originally do not support, but support now. | NA | Yes | Yes | Yes | No | Step 1 to step 4 |
| QUO | CORBA specific framework | Any domain | Yes | Yes | NA | Yes | Yes | Yes | No | Step 1 to step 4 |
| SLAng | XML Language | Originally for Internet DS environment | NA | Yes | No But based on behavior of SLA parties | NA | Yes | Yes | Yes | Step 1 to Step 4 |

### 3.2.4. Monitor - SLA Violation

Monitoring infrastructures are used to measure the difference between the pre-agreed and actual service provision between parties (Rana et. al. 2008). There are three types of monitoring infrastructures, which are trusted third party (TTP), trusted module on the provider side, and trusted module on the client side. Nowadays, TTP provides most of the functionalities for monitoring in most typical situations to detect SLA violation.

### 3.2.5. Terminate - SLA

There are two scenarios in which an SLA may be terminated. The first is termination due to normal time out. The second one is termination because any party violated its contract terms. Normally, in Clouds, this step is conducted by customers and termination typically is caused by normal time out or the provider's SLA violation. Sometimes, providers also terminate SLAs depending on the task priorities. If the reason for SLA termination is violation, then the 'Enforce Penalties for SLA Violation' step of the SLA lifecycle has to be applied. Usually this step is performed manually..

### 3.2.6. Enforce Penalties for SLA Violation

A penalty clause can be applied to the party who violates SLA terms. First is a direct financial compensation being negotiated and agreed between parties. Second is a decrease in price along with the extra compensation for any subsequent interaction. In other words, this option is according to the value of loss caused by the violation. In this case, TTP is usually used as a mediator. The workflow for this option is that clients transfer their deposit, bond, and any other fees into the Third Party's account, and then if the SLOs have been met, the money is paid to provider via TTP. Otherwise, the TTP returns the amount of fees back to the consumer as compensation for SLA violations. The SLA violation has two indirect side impacts on providers. The first is that consumers will use less service from the provider in the future. The second is that provider' reputation decreases and it affects other clients' willingness to choose this provider subsequently. The major indirect influence on consumer is that future request will be rejected due to bad credit record.

A major issue, in the above discussion, is the variety of laws enforced in different countries. This problem can be solved by a '*choice of law clause*', which indicates explicitly which country's laws are applied when a conflict occurs between parties. '*Legal templates*' (Dinesh, 2004) can be used to refine these clauses (Rana et. al. 2008).

## 4. SLA USE CASES IN UTILITY COMPUTING SYSTEMS

Utility computing provides access to on-demand delivery of IT capabilities to the consumer according to cost-effective pricing schema. Typically, a resource in a Data Center is idle during 85% of time (Yeo et. al. 2008). Utility computing provides a way for enterprises to lease this 85% of idle resource or to use outsourcing to pay for resources according to their usage. Two approaches of utility computing that achieve above goals are Grid and Cloud. In the remaining part of this section, we present use cases in Grid and Cloud computing environments.

### 4.1. SLA in Grid Computing Systems

In this section we introduce the definition of Grid computing, and some recent significant Grid computing projects that have focused on SLAs and enabled them in their frameworks.

According to Buyya et. al. (2009) "*A Grid is a type of parallel and distributed system that enables the sharing, selection, and aggregation of geographically distributed 'autonomous' resources dynamically at runtime depending on their availability, capability, performance, cost, and users' quality-of-service re-* *quirements.*" Grid computing is a paradigm of utility computing, typically used for access to scientific resources, even though it has been also used in the industry as well.

SLA has been adopted in Grid computing, and many Grid projects are SLA oriented. We classify them into three categories, which are SLA for business collaboration, SLA for risk assessment, and SLA renegotiation supporting dynamic changes.

**SLA for Business Collaboration:** GRIA (The GRIA Project) is a service-oriented infrastructure designed to support B2B collaborations across organizational boundaries by providing services. The framework includes a service manager with the ability to identify the available resources (e.g. CPUs and applications), assign portions of the resources to consumers by SLAs, and charge for resource usage. Furthermore, a monitoring service is responsible for monitoring the activity of services with respect to agreed SLOs.

The BREIN consortium (The BREIN Project, 2006-2009) defines a business framework prototype for electronic business collaborations. Some capabilities of this framework prototype include Service Discovery with respect to SLA capabilities, SLA negotiation in a single-round phase, system monitoring and evaluation, and SLA evaluation with respect to the agreed SLA. The WSLA/WS-Agreement specifications are suggested for SLAs management. The project focuses on dynamic SLAs. This initiative shows that the industry is demonstrating their interest in SLA management.

In the work of Joita et. al. (2005), WS-Agreement specification is used as a basis to conduct negotiation between two parties. An agent-based infrastructure takes care of the agreement offer made by the requesting party. In this scenario, many one-to-one negotiations are considered in order to find the service that matches the offer best.

**Risk Assessment:** The AssessGrid (Battre et. al. 2007) project focuses on risk management and assessment in Grid. It aims at providing service providers with risk assessment tools, which help them to make decisions on the suitable SLA offer by assigning, mapping, and associating the risk of failure to penalty fees. Similarly, end-users get knowledge about the risk of an SLA violation by a resource provider that helps them to make appropriate decisions regarding acceptable costs and penalty fees. A broker is the matchmaker between end-users and providers. WS-Agreement-Negotiation protocol is responsible for negotiating SLAs with external contractors.

**SLA renegotiation supporting dynamic changes:** Frankova et. al. (2006) propose an extension of WS-Agreement allowing a run-time SLA renegotiation. Some modifications are proposed in the 'Guarantee-Term' section of the agreement schema and a new section is added to define possible negotiations, to be agreed by parties before the offer is submitted. The limitation is that it does not support run-time renegotiation to adapt dynamic operational and environmental changes, because after the agreement's acceptance, there is no interaction between the provider and the consumer. Sakellariou et. al. (2005) specify the guarantee terms of an agreement as variable values rather than fixed values. This work aims at minimizing the number of re-negotiations to reach consensus with agreement terms. BabelNet, is a Protocol Description Language for automated SLA negotiation, has been proposed (Hudert et. al. 2009) to handle multiple-dimensional auctions.

## 4.2. SLA in Cloud Computing

Cloud computing is a paradigm of service oriented utility computing. In this section we introduce a definition of cloud computing and SLA use cases in industry and academia. Finally, we compare SLA usage difference between Cloud computing and traditional web services.

### 4.2.1. Cloud Computing

Based on the observation of the essence of what Clouds are promising to be, Buyya et. al. (2009) propose the following definition: "*A Cloud is a type of parallel and distributed system consisting of a collection of inter-connected and virtualized computers that are dynamically provisioned and presented as one or more unified computing resource(s) based on service-level agreements established through negotiation between the service provider and consumer*". Hence, Clouds fit well into the definition of utility computing.

*Figure 6* shows the layered design of Cloud computing architecture. Physical Cloud resources along with core middleware capabilities form the bottom layer needed for delivering IaaS. The user-level middleware aims at providing PaaS capabilities. The top layer focuses on application services (SaaS) by making use of services provided by the lower layer services. PaaS/SaaS services are often provided by 3rd party service providers, who are different from IaaS providers. (Buyya et. al. 2009)

**User-Level Applications:** this layer includes the software applications, such as social computing applications and enterprise applications, which will be deployed by PaaS providers renting resources from IaaS providers.

**User-LevelMiddlewire:** Cloud programming environments and tools are included in this layer facilitate creation of applications and their mapping to resources using Core Middleware Layer services.

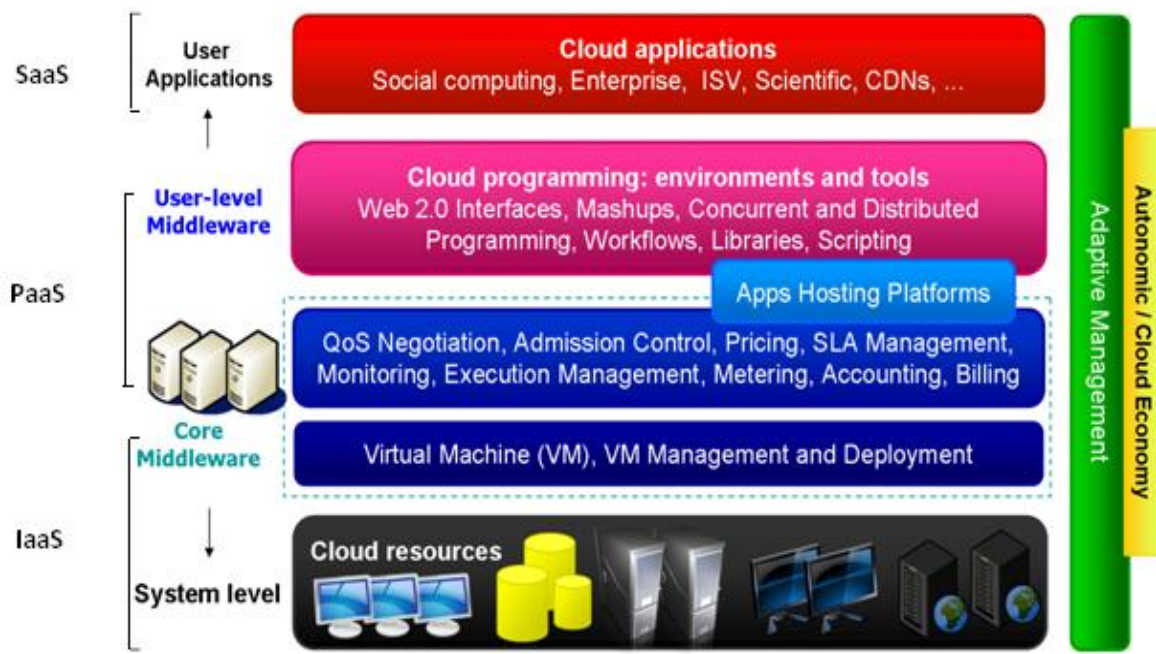

*Figure 6 : Layered Cloud computing architecture.* (Buyya et. al 2009)

**Core Middleware:** this layer provides runtime environment enabling capabilities to application services built using User-Level Middleware. Dynamic SLA management, Accounting, Monitoring and Billing are examples of core services in this layer. The commercial examples for this layer are Google App Engine and Aneka.

**System Level:** physical resources including physical machines and virtual machines sit in this layer. These resources are transparently managed by higher level virtualization services and toolkits that allow sharing of their capacity among virtual instances of servers.

### 4.2.2. Use Cases

In this section, we present industry and academic use cases in Cloud computing environments.

**Industry Use Cases**: In this section, we present how Cloud providers implement SLA. Important parameters are summarised in Table 4. All elements in Table 4, obtained from formal published SLA documents of AmazonEC2 and S3 (IaaS provider), and Windows Azure1 Compute and Storage (IaaS/PaaS provider).

A Characterization of systems studied following the six steps of SLA lifecycle model is summarized in *Table 5*. From the users' perspective, the process of activating SLA lifecycle with Amazon and Microsoft is simple because the SLA has been pre-defined by the provider. According to SLA lifecycle, the first step is to find the service providers according to users' requirements. For example, users find the provider via searching on the Internet, and then explore the providers' web site for collecting further information. Most Cloud service providers offer pre-defined SLA documents. In this case, the second step and third step are pre-defined and always be entwined together. The check for SLA violation monitoring can be done by third party tools, such as Cloudwatch, Cloudstatus, Monitis, and Nimsoft. Developers are able to develop their own monitoring systems by using these tools.

For what concerns the termination of an SLA we can consider IaaS services as a reference example. In this case three scenarios may occur. The normal termination of an SLA is constituted by the release of Cloud release of Cloud resources by the user. An SLA can also be actively terminated by a provider if the resource usage lasts beyond the predefined expiry time. A termination with penalty may occur in case the provider is unable to provide resources according to the expected Quality of Service. The last step of SLA lifecycle will be invoked if any party violates contract terms. Currently most of the service providers give service credits to customer if they violate SLA.

*Table 4: SLA Use Cases of the most famous Cloud Provider and related characteristics in SLAs*

| Cloud Provider Name | Service Commitment | Effective Date | Monthly Uptime Percentage (MUP)% | Service Credits Percentage (%) |
|---|---|---|---|---|
| **Amazon AWS EC2** | "AWS use commercially reasonable efforts to make Amazon EC2 available with an Annual Uptime Percentage of at least 99.95% during the Service Year. In the event Amazon EC2 does not meet the Annual Uptime Percentage commitment, you will be eligible to receive a Service Credit "(AWS EC2 Service Level Agreement). | October 23, 2008 | $MUP_1$<99.95% | **10%** |
| **Amazon AWS S3** | "AWS use commercially reasonable efforts to make Amazon S3 available with a Monthly Uptime Percentage (defined below) of at least 99.9% during any monthly billing cycle (the "Service Commitment"). In the event Amazon S3 does not meet the Service Commitment, you will be eligible to receive a Service Credit "(AWS S3 Service Level Agreement). | October 1, 2007 | 99%=<MUP<99.9% MUP<99 | **10%** **25%** |
| **Windows Azure Compute** | "Windows Azure has separate SLA's for compute and storage. For compute, we guarantee that when you deploy two or more role instances in different fault and upgrade domains your Internet facing roles will have external connectivity at least 99.95% of the time. Additionally, we will monitor all of your individual role instances and guarantee that 99.9% of the time we will detect within two minutes when a role instance's process is not running and initiate corrective action." (Windows Azure Service Level Agreement) | NA | <99.95% <99% | **10%** **25%** |
| **Windows Azure Storage** | | NA | <99.9% <99.5% | **10%** **25%** |

1.The formula used to calculate Monthly Connectivity Uptime Percentage (MCUP) is depending on Maximum Connectivity Minutest (MCM), Connectivity Downtime (CD) and Maximum Connectivity Minutest (MCM). The equation is given as follows $MCUP = (MCM - CD) \div MCM$    *Source: Windows Azure Service Level Agreement*

| Cloud Service Provider | Service Type | Step 1: Discover-Service Provider | Step 2: Define-SLA | Step 3: Establish-Agreement | Step 4: Monitor-SLA Violation | Step 5: Terminate-SLA | Step 6: Enforce Penalties for SLA Violation |
|---|---|---|---|---|---|---|---|
| **Amazon EC2** | IaaS (Computing) | Discover manually (e.g. via web site) | Pre-defined SLA terms and QoS parameters | Pre-defined SLA document by provider | Can use third party monitor systems (e.g. Cloud-Watch) | By user, or provider program-matically or manual-ly | Service Credit given by provider |
| **Amazon S3** | IaaS (Storage) | Discover manually | Pre-defined SLA terms and QoS parameters | Pre-defined SLA document by provider | Can use third party monitor systems (e.g. CloudSta-tus) | By user, or provider program-matically or manual-ly | Service Credit given by provider |
| **Microsoft Azure Compute** | PaaS | Discover manually (e.g. via web site) | Pre-defined SLA terms and QoS parameters | Pre-defined SLA document by provider | Can use third party monitor systems (e.g. Moni-tis) | By user, or provider program-matically or manual-ly | Service Credit given by provider |
| **Microsoft Azure Storage** | PaaS | Discover manually | Pre-defined SLA terms and QoS parameters | Pre-defined SLA document by provider | Can use third party monitor systems (e.g. Moni-tis) | By user, or provider program-matically or manual-ly | Service Credit given by provider |

**Academic Use Cases:** In this section, we present SLA-Oriented projects and algorithms as academy use cases.

**SLA-Oriented Resource Allocation for Data Centers and Cloud Computing Systems**: The Cloud Computing and Distributed Systems (CLOUDS) Laboratory, at the University of Melbourne has proposed the use of market-based resource management to support utility-based resource management for cluster computing (Yeo C. S. et. al. 2005) (Yeo C. S. et. al. 2007). The initial work successfully demonstrated that market-based resource allocation strategies are able to deliver better utility for users than traditional system-centric strategies. However, early research focused on satisfying only two static Quality of Service (QoS) parameters: the deadline for completing a service request and the budget that the consumer is willing to pay for completing the request before the deadline. In the commercial computing environment, there are other critical QoS parameters to consider in a service request, such as reliability and trust/security. In particular, QoS requirements cannot be static and need to be dynamically updated over time due to continuing changes in business operations and operating environments.

**SLA@SOI**: A European Union funded Framework 7 research project, SLA@SOI (SLA@SOI project), is researching aspects of multi-level, multi-provider SLAs within service-oriented infrastructure and cloud computing. Currently, this project aims to build an ad-hoc architecture and integration approach for a basic SLA management framework. It provides a major milestone for the further evolution towards a ser-vice-oriented economy, where IT-based services can be flexibly traded as economic goods, i.e. under well defined and dependable conditions and with clearly associated costs. SLA@SOI provides two major benefits to the provisioning of services. First, service predictability and dependability means that the quality characteristics of service can be predicted and enforced at run-time. Second, automation means that the whole process of negotiating SLAs and provisioning, delivery and monitoring of services can be automated allowing highly dynamic and scalable service consumption.

**SLA based Management and Scheduling:** Lee et. al. (2010) propose profit-driven SLA based scheduling algorithms in Clouds to maximize the profit for service providers. The application model used in this work can be classified as SaaS and PaaS. The service types supported by their algorithm are dependent services, which mean one sub-service can not start until the prerequisite services are completed. However, their work does not support multiple providers and full simulation configuration is not available. We recommend possible future research direction is SLA management with multiple providers, since it is required for emerging research in InterCloud.We define InterCloud as multiple Cloud providers with peer agreement to support collaborative activities.

## 4.2.4 SLA related difference between Cloud and Web Service

In this section we compare the differences between SLAs applied in cloud computing and in traditional web services as follows:

**QoS Parameters**: Most web services focus on parameters such as response time, SLA violation rate for the task, reliability, availability, levels of user differentiation, and cost of service. In Cloud computing more QoS parameters than traditional web services need to be considered, for example, energy related QoS, Security related QoS, Privacy related QoS, trust related QoS. More than 20 QoS parameters are defined by the SMI (Service Management Index) consortium to be used in the industry and academia as standard benchmark.

**Automation:** The whole process of SLA negotiation and provisioning, service delivery and monitoring needs to be automated for highly dynamic and scalable service consumption. Researchers in traditional web services explored this topic, for example, Jin L.J et. al. (2002) proposed a model for SLA analysis of Web Services. Nevertheless, SLA automation is a rapidly growing area in Cloud computing. In fact there are some research projects starting to focus on it, such as CLOUDS Lab at the University of Melbourne and SLA@SOI.

**Resource Allocation**: SLA oriented resource allocation in Cloud computing is possible different from allocation in traditional web services, because web services have a Universal Description Discovery and Integration (UDDI) for advertising and discovering between web services. However, in Clouds, resources are allocated and distributed globally without central directory, so the strategy and architecture for SLA based resource allocation in such environment are different from traditional web services.

## 5. ONGOING WORKS

SLA management must provide ways for reliable provisioning of services, monitoring of SLA violations and detection of any potential performance decrease during service execution (Kuo et. al. 2006) (Marilly et. al. 2002). The goal of SLA management is to establish a scalable and automatic SLA management framework that can adapt to dynamic environmental changes by considering multiple QoS parameters. In addition, an SLA has to be suitable for multiple domains with heterogeneous resources. Some of the research are works towards to this direction. The VIRD architecture is a three-level hierarchy focused on scalability. Wurman et. al. (1998) state a set of auction parameters and price-based negotiation platform. Nevertheless, this solution only supports one-dimensional auction, thus could not handle multiple-dimensional auctions, which are important in utility computing environments. Recently, BabelNet handles multiple-dimensional auctions.

Nevertheless, somehow consumers still need to be involved in the management process to certain extent. Moreover, multiple QoS parameters have been investigated by CLOUDS Lab's initial work. Whilst that work only focused on the most common QoS parameters (price and deadline), there are other critical QoS parameters that should be considered in a service request, such as reliability and trust/security. In particular, QoS parameters are must be updated dynamically over time due to continuing changes in business operating environments. Thus, multiple QoS parameters should be investigated in the future research work.

More specifically, there are some open challenges for SLA management. First and foremost, different SLA negotiation protocols and processes constrain the negotiation for establishing SLAs, the modification of an implemented SLA, and SLA negotiation between distinct administrative domains. Second, The SLA has to be established between providers and consumers from different end-to-end viewpoint. For example, if the system service has been outsourced from one provider to another, there should be SLA agreement between them as well. Third, admission control policies need to be defined, because decision on which user request to accept affects the performance, profit, and reputation of the resource provider. Moreover, the resource allocation management has to be considered carefully, because it addresses which resource is best suitable for currently admitted requests from both parties' point of view. Some termination related problems are management of QoS metrics, different parties use different parameters, and the failure to manage becomes an issue especially for the automatic handling, such as cause analysis, automatic problem resolution. We can also mention, performance forecast management is another open question in utility computing environments because it enables the recommendation for performance improvement.

## 6. SUMMARY

This chapter presented the literature survey, issues and solutions of SLA management in utility computing systems and how SLAs have been used in these systems. An SLA is a formal contract between service providers and consumers to guarantee that the service quality is delivered to satisfy pre-agreed consumers' expectations. SLA management is important in utility computing systems because it helps to improve the customer satisfaction level and to define clear relationship between parties. In this chapter, we have summarised the main fundamental concepts of SLA and analyzed two types of SLA lifecycle. One is the three phase high level lifecycle, which includes creation phase, operation phase and removal phase; the other is more specific lifecycle including six steps, which are 'discover-service provider', 'define-SLA elements', 'establish-agreement', 'monitor-SLA violation', 'terminate-SLA' and 'SLA violation control'. The second type of lifecycle is more comprehensive, and introduces the characterization of SLA violation that is a foundation in utility computing environments where services are consumed on a pay-as-you-go basis.

The analysis carried out in this book chapter has identified four major goals in case of SLA-oriented utility computing. First, supporting customer-driven service management based on customer profiles and requested service requirements. Second, defining computational risk management tactics to identify and manage risks involved in the execution of applications with regards to service requirements and customer needs. Third, deriving appropriate market-based resource management strategies encompassing both customer-driven service management and computational risk management to sustain SLA-oriented resource allocation. Fourth, incorporating autonomic resource management models and self-manage changes in service requirements to satisfy both new service demands and meet existing service obligations.

To achieve these goals, we discussed the main challenges and solutions of SLA implementation and management in utility computing environments by following the steps of SLA lifecycle. In the 'discover-service provider', the main issues are scalability, dynamic changes, heterogeneity, and autonomous administration. Some architectures and algorithms have been proposed to cope with them, such as MDS architecture and the VIRD architecture. Effective negotiation protocols and processes are main challenges for the 'define-SLA' and 'establish- agreement' steps, because two parties need to negotiate before they agree on the terms that have to be included in SLAs. SLA frameworks and languages are used as solutions. Currently the most widely used languages are WSLA and WS-Agreement. However, there are not many effective solutions for the automatic negotiation. Thus, the automatic negotiation is still an open issue. Regarding the 'monitor SLA violation', which party should be responsible for the monitoring process is a debate issue. The most popular solution for this problem is using Third Party (TTP) who provides most of functionalities for monitoring a service in most typical situations to detect SLA violations. The main issues for the last two steps 'terminate SLA' and 'enforce penalties for SLA violation', are automatic failure management, such as cause analysis, penalty clauses invocation, and automatic failure resolution. Some penalty strategies have been presented. However, automatic problem resolution and cause analysis are still open challenges and more investigation is needed in the future.

In conclusion, SLA in utility computing systems is a rapidly moving target although some works have been explored in the past. Therefore, there are still some open challenges such as scalability, dynamic environmental changes, heterogeneity, SLA management automation, multiple QoS parameters, and SLA suitable for cross domains need to be explored in future research.

## ACKNOWLEDGEMENTS

The authors would like to acknowledge all researchers of works described in this book chapter and thank them for their outstanding work. We also thank for Yoganathan Sivaram, Christian Vecchiola, Saurabh Kumar Garg, Rodrigo Calheirós, William Voorsluys, Tong Zou, Shanshan Wu and Daryl de Penha for their comments to improve the quality of this book chapter.

## ADDITIONAL READING SECTION

Bastian Koller, Eduardo Oliveros, Alfonso Sánchez-Macián, Service Level Agreements in the Grid Environment, Market Oriented Grid and Utility Computing, R. Buyya and K. Bubendorfer (eds), ISBN: 978-0470287682, Wiley Press, Hoboken, New Jersey, USA, Oct. 2009.

Paul McKee, Steve Taylor, Mike Surridge, and Richard Lowe, SLAs, Negotiation and Potential Problems, Market Oriented Grid and Utility Computing, R. Buyya and K. Bubendorfer (eds), ISBN: 978-0470287682, Wiley Press, Hoboken, New Jersey, USA, Oct. 2009.

Marco A. S. Netto, Kris Bubendorfer, and Rajkumar Buyya, SLA-based Advance Reservations with Flexible and Adaptive Time QoS Parameters, Proceedings of the 5th International Conference on Service-Oriented Computing (ICSOC 2007, LNCS Volume 4749, Springer-Verlag Press, Berlin, Germany), Sept. 17-20, 2007, Vienna, Austria.

Rajiv Ranjan, Aaron Harwood, and Rajkumar Buyya, SLA-Based Coordinated Superscheduling Scheme for Computational Grids, Proceedings of the 8th IEEE International Conference on Cluster Computing (Cluster 2006, IEEE CS Press, Los Alamitos, CA, USA), Sept. 27-30, 2006, Barcelona, Spain.

James Broberg, Srikumar Venugopal, Rajkumar Buyya, Market-oriented Grids and Utility Computing: The state-of-the-art and future directions, Journal of Grid Computing, Volume 6, Number 3, Pages: 255-276, ISSN: 1570-7873, Springer Verlag, Germany, Sept. 2008.

Rajkumar Buyya and Srikumar Venugopal, The Gridbus Toolkit for Service Oriented Grid and Utility Computing: An Overview and Status Report, Proceedings of the First IEEE International Workshop on Grid Economics and Business Models (GECON 2004, April 23, 2004, Seoul, Korea), 19-36pp, ISBN 0-7803-8525-X, IEEE Press, New Jersey, USA.

Jordi Guitart, Mario Macías, Omer Rana, Philipp Wieder, Ramin Yahyapour, and Wolfgang Ziegler, SLA-based Resource Management and Allocation, Market Oriented Grid and Utility Computing, R. Buyya and K. Bubendorfer (eds), ISBN: 978-0470287682, Wiley Press, Hoboken, New Jersey, USA, Oct. 2009.

Rajkumar Buyya, Srikumar Venugopal, Rajiv Ranjan, and Chee Shin Yeo, The Gridbus Middleware for Market-Oriented Computing, Market Oriented Grid and Utility Computing, R. Buyya and K. Bubendorfer (eds), ISBN: 978-0470287682, Wiley Press, Hoboken, New Jersey, USA, Oct. 2009.

## AUTHORS PROFILE

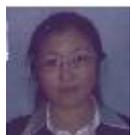

Linlin Wu is a PhD candidate under the supervision of Professor Rajkumar Buyya in the CLOUDS Laboratory at the University of Melbourne, Australia. She received Master of Information Technology from the University of Melbourne and then worked for CA (Computer Associates) Pty Ltd) as Quality Assurance Engineer. Then she joined National Australia Bank (NAB) as a Knowledge Optimization Officer. Here in Melbourne University, she has been awarded with APA scholarship supporting PhD studies. She received the Best Paper Award from AINA 2010 conference for her first publication. Her current research interests include: Service Level Agreement, QoS measurement, Resource Allocation, and Market-oriented Cloud computing.

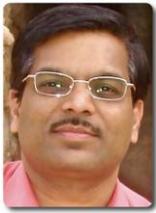 Dr. Rajkumar Buyya is Professor of Computer Science and Software Engineering; and Director of the Cloud Computing and Distributed Systems (CLOUDS) Laboratory at the University of Melbourne, Australia. He is also serving as the founding CEO of Manjrasoft Pty Ltd., a spin-off company of the University, commercializing its innovations in Grid and Cloud Computing. He has authored and published over 300 research papers and four text books. The books on emerging topics that Dr. Buyya edited include, High Performance Cluster Computing (Prentice Hall, USA, 1999), Content Delivery Networks (Springer, Germany, 2008), Market-Oriented Grid and Utility Computing (Wiley, USA, 2009), and Cloud Computing (Wiley, USA, 2011). He is one of the highly cited authors in computer science and software engineering worldwide. Software technologies for Grid and Cloud computing developed under Dr. Buyya's leadership have gained rapid acceptance and are in use at several academic institutions and commercial enterprises in 40 countries around the world. Dr. Buyya has led the establishment and development of key community activities, including serving as foundation Chair of the IEEE Technical Committee on Scalable Computing and four IEEE conferences (CCGrid, Cluster, Grid, and e-Science). He has presented over 250 invited talks on his vision on IT Futures and advanced computing technologies at international conferences and institutions in Asia, Australia, Europe, North America, and South America. These contributions and international research leadership of Dr. Buyya are recognized through the award of "2009 IEEE Medal for Excellence in Scalable Computing" from the IEEE Computer Society, USA. Manjrasoft's Aneka technology for Cloud Computing developed under his leadership has received "2010 Asia Pacific Frost & Sullivan New Product Innovation Award".